# Large-Scale 3D Printing - Market Analysis

Razan Abdelazim Idris Alzain

Campus Ring 7, 28759 Bremen, Germany

# Table of Contents

# List of figures

# List of tables

# List of abbreviations

| | |
|---|---|
| 3D | Three-dimensional |
| CAD | Computer-aided Design |
| CAM | Computer-aided Manufacturing |
| FDM | Fused Deposition Modeling |
| FFF | Fused Filament Fabrication |
| ABS | Acrylonitrile Butadiene Styrene |
| PLA | Polylactic Acid |
| SLS | Selective Laser Sintering |
| EBM | Electron Beam Melting |
| BAAM | Big Area Additive Machine |
| DDD | 3D Systems Corp. |
| PRLB | Proto Labs Inc. |
| FARO | FARO Technologies Inc. |
| MTLS | Materialise NV |
| XONE | The ExOne Co. |
| API | Application Programming Interface |
| AM | Additive Manufacturing |
| BAAM | Big Area Additive Machine |

# 1   Introduction

## 1.1   Problem of the paper

One of the main problems brought by large-scale 3D printers is the lack of standardization of machines and the potential of low-quality products. If a company invests in an inexpensive 3D printer then the risk of a bad quality product increases. On the other hand, a high-end Large-scale 3D printer would cost millions of dollars to produce a trustworthy result. Regardless, the traditional manufacturing route will always be preferred by production companies. Many different 3D printers produce very different products, making there a lack of universal standards in 3D printing technologies. Therefore, manufacturers compare their products with other manufacturers' methods worrying that they would vary in terms of quality, strength, and reliability. This causes a continuous wariness in the 3D printing technology, making companies always judge the risks compared to the benefits.

Another problem that comes with large-scale 3D printers is the short product lifespan. Yes, having the ability to print a large number of spare parts on-demand can help to prolong product warranties while also being significantly more ecologically friendly, but it is not a smart move. Unfortunately, many businesses rely on a business plan that revolves around low-quality items and product rotation. 3D printing is a concern because it reduces product obsolescence. Firms will need to develop new business models that ensure quality and product lifespans and do not rely on continually creating new items. Furthermore, 3D printing allows customers to create their own spare parts, which is bad for businesses[1].

As mentioned above, 3D printing companies may be able to begin printing their own product parts. This is exacerbated by the fact that smaller enterprises may begin printing product components and entire goods, stealing the intellectual property of larger corporations.

---

[1] Cf (*The big challenges of 3D printing* 2019)

## 1.2    Aims of the paper

The aim of this research is to get a better understanding of the future of large-scale 3D printing. By developing the market analysis, it will be clear whether large-scale 3D printing is becoming more of a preferred way of printing custom-made parts for production companies. Companies can then choose whether to change their ways, for a more profitable less costly method, or stay on the route they are on. By getting deep into this topic, a new world of technology is then being discovered and familiarized. With a mix of theoretical and practical relevance, a complete coverage could be made on large-scale 3D printing. This paper could then cover all aspects of this topic, and the reader could then make their own judgment if large-scale 3D printing would be the best option.

## 1.3    Course of research

The way that is planned on approaching this research is by introducing the topic and familiarizing the reader with Large-scale 3D printers. Then, to move into the literature review, where some reports and academic publications will be covered on large-scale 3D printing. Then proceed with mentioning the benefits and risks that come with 3D printing. A comparison will also be done between small-scale and large-scale 3D printing.

A market analysis is then made for large-scale 3D printing. Deep research will be done to see what the market wants and expects from them. Also a few mentions about the companies that create large-scale 3D printers and the companies that rent/buy/use them. Also make a cost analysis, of the revenue and expenses that come with large-scale printers, to better understand their market value of them.

Afterward, the discussion follows and a limitation section. The limitation section will include features that prevent the large-scale 3D printer from achieving a higher market value. Within this section, a paragraph will be dedicated to the research gap that comes with the internet search limitation. Then, the research will move to the conclusion and then references. As previously mentioned, this research will include a literature review, which will explain a big part of this research's aim. Then followed by the market analysis, which will be made, by providing an overall industry market than focusing on the targeted market. Also by distinguishing the customer characteristics, we can understand the exact appeal of large-scale 3D printers.

An important step in the market analysis is to compare large-scale 3D printers with their competitors. To see which dominates the market now, and why. Later on, all the data must be gathered and then analyzed.

# 2   3D Printing

What is 3D printing? To answer that question in the most basic way, it is said to be a manufacturing technique where the material is layered onto each other layer by layer to build a three-dimensional part/item. With the help of a computer-aided design (CAD) or computer-aided manufacturing (CAM), programmers can create and convert the digital files containing the 3D data into physical objects[2].

For decades, the car industry has been exploring the possibility of 3D printing. 3D printing is particularly beneficial for quick prototyping and has demonstrated the ability to dramatically reduce design and lead times on new automobile models. The manufacturing process has also been improved by 3D printing in the sector. Specialized jigs, fixtures, and another tooling that would be required for a single automobile part, particularly for high-performance machines, used to necessitate a slew of custom tools, adding expense and complicating the process as a whole[3]. Custom jigs and other low-volume items may be made directly for the production line using 3D printing. Manufacturers can reduce lead times by up to 90% and reduce risk by integrating 3D printing technologies. The manufacturing process as a whole becomes more efficient and lucrative by simplifying with in-house production.

3D printing is causing a design revolution in the jewelry industry. It used to be difficult to create 3D printed items that looked and felt like conventional handmade and cast jewelry. However, with the most recent wave of advancements in specialized high-end 3D modeling tools and more printable materials available, an increasing number of jewelry designers are now preferring to 3D model and print their creations over conventional handcrafted methods.

Weight reduction is one of the key ways that 3D printing has helped the aircraft industry to save money. The lower volume of components required in 3D printed construction of a part results in lighter overall parts—this seemingly minor change in production positively affects an aircraft's payload, emissions, and fuel consumption, as well as speed and safety, while significantly reducing production waste. The approach, like in many other sectors, enables the manufacturing of components that are just too complicated for traditional ways to manage.

---

[2] Cf *(3D Printing: What You Need to Know, 2022)*
[3] Cf *(25 (Unexpected) 3D Printing Use Cases, 2022)*

## 2.1    Type of Technology

Different technologies are used by 3D printers. The most well-known is fused deposition modeling (FDM), often known as fused filament manufacturing (FFF). This is manufactured from acrylonitrile butadiene styrene (ABS), polylactic acid (PLA), or another thermoplastic filament. They are then melted together and discharged in layers through a heated extrusion tube. The first 3D printers to hit the market, created by Stratasys in the mid-1990s, utilized FDM, as do most 3D printers aimed at consumers, enthusiasts, and schools today. FDM is utilized in 3D printed structures by extruding clay or concrete, 3D printed desserts by extruding chocolate, 3D printed organs by extruding living cells in a bio gel, and so on. If anything can be extruded, it can almost certainly be 3D printed.

Stereolithography is another 3D printing process. In it, a UV laser is shone into a vat of ultraviolet-sensitive photopolymer, tracing the surface of the item to be made. The polymer hardens wherever the beam comes into contact with it, and the beam "prints" the item layer by layer according to the instructions in the CAD or CAM file from which it is working. A variant of this is a digital light projector (DLP) 3D printing. A liquid polymer is exposed to light from a digital light processing projector in this approach. This hardens the polymer layer by layer until the item is completed, at which point the residual liquid polymer is drained away.

SLA has the distinction of being the world's first 3D printing technology. Chuck Hull devised stereolithography in 1986, filing a patent on the technology and establishing 3D Systems to market it.[4] An SLA printer employs mirrors known as galvanometers or galvos, one on the X-axis and one on the Y-axis. These galvos fire a laser beam over a vat of resin, selectively curing and hardening a cross-section of the item inside the construction area, layer by layer.

---

[4] Cf *(All3DP, 2018)*

Multi-jet modeling is a 3D printing technology that sprays a colored, glue-like binder over consecutive layers of powder where the item is to be produced, similar to an inkjet printer. This is one of the quickest ways, as well as one of the few that offers multicolor printing. It is feasible to alter a regular inkjet printer so that it can print using materials other than ink. Enterprising do-it-yourselfers have developed or modified print heads, primarily piezoelectric heads, to operate with a variety of materials—in some cases printing the print heads themselves on other 3D printers! MicroFab Technologies, for example, sells 3D-capable print heads (as well as complete printing systems).

There is also selective laser sintering (SLS), which is a high-powered laser used to fuse particles of plastic, metal, ceramic, or glass. On the other hand, Electron beam melting (EBM) melts metal powder layer by layer using an electron beam. Titanium is frequently combined with EBM to create medical implants and aeronautical components.

## 2.2   Benefits

Designers may use 3D printing to swiftly convert thoughts into 3D models or prototypes (a.k.a. "rapid prototyping") and execute rapid design revisions. It enables producers to make things on demand rather than in huge batches, which improves inventory management and reduces warehousing space. People in faraway areas may now create items that would otherwise be unavailable to them.

In practice, 3D printing may save money and material over subtractive processes since very little raw material is lost. It also promises to transform the nature of production by someday allowing people to download data for printing even complicated 3D products, such as electrical gadgets, in their own homes.

Another advantage of 3D printing is that it enables the creation and production of more complicated designs than traditional manufacturing procedures. Traditional techniques have design constraints that are no longer applicable with the usage of 3D printing. 3D printing can also produce parts in a matter of hours, which expedites the prototype process. This allows each step to be completed more quickly. When compared to machining prototypes, 3D printing is less expensive and faster at generating components since the part may be produced in hours, allowing each design alteration to be performed at a much faster rate[5].

Another advantage of print on demand is that, unlike traditional printing procedures, it does not need a large amount of storage space for inventories. This saves space and money by eliminating the need to print in bulk until absolutely essential. The 3D design files are all maintained in a virtual library and may be searched and printed as needed since they are produced using a 3D model as a CAD or STL file. Design modifications may be made at a low cost by modifying individual files rather than discarding out-of-date items and investing in new equipment.

3D printing is being utilized in the medical field to save lives by printing human organs such as livers, kidneys, and hearts. Further advancements and applications are being explored in the healthcare sector, which will provide some of the most significant benefits of adopting the technology.

## 2.3    Disadvantages

While 3D printing can make products from a variety of polymers and metals, the accessible raw materials are not exhaustive. This is because not all metals or polymers can be thermally regulated sufficiently to allow 3D printing. Furthermore, many of these printing materials are not recyclable, and just a handful are food safe.

---

[5] Cf *(What are the Advantages and Disadvantages of 3D Printing?, 2022)*

Currently, 3D printers feature limited print chambers that limit the size of items that can be manufactured. Anything larger will have to be printed in separate sections and then assembled afterward. Because the printer needs to produce more components before manual labor is employed to connect the parts together, this can raise costs and time for bigger parts.

Although big pieces, as previously stated, require post-processing, most 3D printed items require some type of cleaning up to remove support material from the build and smooth the surface to obtain the desired quality. Waterjetting, sanding, a chemical soak and rinse, air or heat drying, assembling, and other procedures are used for post-processing. The quantity of post-processing required is determined by a variety of factors, including the size of the item being produced, the intended application, and the type of 3D printing technique utilized in manufacturing. As a result, while 3D printing allows for the rapid creation of parts, post-processing might impede the manufacturing process.

As 3D printing becomes more popular and accessible, there is a higher potential that individuals may make false and counterfeit items that will be nearly difficult to distinguish. This has obvious implications for both copyright and quality control.

Another potential issue with 3D printing is directly tied to the type of machine or method utilized; certain printers have lesser tolerances, which means that finished items may deviate from the original design. This can be corrected in post-production, but keep in mind that it will increase the time and expense of production.

# 3    Large-Scale Printers

To be said simply, large-scale 3D printing is the industrial-scale 3D printing of previously molded or machined products. Consider printing a life-size mannequin, a larger-than-life Coke bottle for advertising, a car's whole bumper, or even an airplane wing. Large-scale 3D printing is not just a less expensive alternative to machining; it can also manufacture complicated shapes that would otherwise need several pieces and assembly. This reduces both production time and end-product costs[6].

Large-scale 3D printing is a fantastic alternative for manufacturers when it comes to developing molds and tooling since it reduces lead times and costs compared to traditional production setup while also reducing the constraints associated with traditional design.

## 3.1    Background

Oak Ridge National Labs, a US Department of Energy research site, pioneered large-scale 3D printing. In terms of sheer scale, they continue to lead the way: the latest Big Area Additive Machine (BAAM) can deposit over 36 kg of material per hour, producing pieces up to 13 feet long, 6.5 feet wide, and 8 feet tall.

BAAM's initial edition was released in 2014 as part of a collaboration with the City of Cincinnati. The most recent iteration, which was used to print a full 3D-printed automobile in 2017, features two hoppers, dryers, and lines to the extruder to enable printing with diverse materials — especially beneficial for creating objects with different qualities on the surface than within.

BAAM and related machines are most typically employed to make big molds — for things like airplane wings or the cladding that many high-rise buildings have. Traditionally, these things would have been made using massive wood molds over which material would be molded or poured. However, making these molds might take months. BAAM permits the creation of a mold in a matter of days.

---

[6] Cf *(What is Large Format 3D Printing and Who Needs It? - XponentialWorks, 2022)*

## 3.2 Today's Large scale printers

The most recent generation of commercial large-format 3D printing technologies, which have already been widely used in a variety of sectors, are not as huge as BAAM – but they are significantly faster and more sophisticated. The Nexa3D NXE400 printer, for example, can print up to 16 liters of component volume per minute at rates of up to 1Z centimeter per minute. By contrast, every other similar 3D printer on the market has six times the speed and 2.5 times the volume of this one.

The Nexa3D printer can also print with strong materials, making it perfect for high-speed printing of functional prototypes, manufacturing tools, full-scale end-use parts, and casting patterns. The printers have intelligent software and integrated sensors, which optimize production performance while also providing thorough diagnostics and continuous monitoring.

Following that, there's no doubting the insaneness of a 600 x 600 x 660 mm construction space for less than $4,000[7]. The Modix Big60 V3 is a true workhorse printer with the tools to make the most of its size, including an E3D Volcano hot end for fast printing. The thinking is handled by a Duet 2 Wi-Fi-enabled mainboard, which is paired with a Duet touchscreen controller, which gives the operator access to macros and other live-print controls. It lacks an enclosure (which costs extra) and requires self-assembly, but it's difficult not to be lured to that build volume.

Finally, we'll take a look at Vivedino (previously Formbot) and its Vivedino Troodon CoreXY 3D printer. The build space of 400 x 400 x 500 mm and Wi-Fi-enabled mainboard are two of the main draws, coupled with the fact that it is based on the Voron project's highly programmable CoreXY 3D printer architecture. At $1,600, it isn't the cheapest, but it has a

---

[7] Cf *(All3DP, 2018)*

substantial build volume and a great feature set. It's also nearly ready to use right out of the box.

**Table 1: List of best Large scale printers in 2022**

| 3D Printer | Build Volume (mm) | Price (USD, approx.) | Check Price (Commissions Earned) |
|---|---|---|---|
| Creality Ender 5 Plus | 350 x 350 x 400 mm | $582 | **Creality3D Official Store** |
| Modix Big-40 | 400 x 400 x 800 mm | $4,900 | **Modix** |
| Modix Big60 V3 | 600 x 600 x 660 mm | $3,900 | **Modix** |
| Tronxy X5SA-500 Pro | 500 x 500 x 600 mm | $820 | **AliExpress** |
| Vivedino Troodon | 400 x 400 x 500 mm | $1,675 | **AliExpress** |
| gCreate gMax 2 | 457 x 457 x 609 mm | $3,995 | **gCreate (no commission)** |

Source: *(All3DP, 2018)*.

## 3.3    Manufacturers

*3D Systems Corp. (DDD)*

3D Systems pioneered 3D printing in 1989 with the development and patenting of their stereolithography technique, which employs ultraviolet lasers to aid in the creation of highly exact parts. DDD expanded on this by innovating additional technologies such as selective laser sintering, multi-jet printing, film-transfer imaging, color jet printing, direct metal printing, and plastic jet printing. 3D Systems operates in three divisions: products, materials, and services. The items category covers tiny desktop and commercial printers that print in plastics and other materials, as well as 3D printers and software[8].

*Proto Labs Inc. (PRLB)*

Proto Labs was established in 1999 with the goal of developing automated solutions for the development of plastic and metal parts utilized in the manufacturing process. The firm grew to create an industrial-grade 3D printing service, allowing developers and engineers to transfer prototypes into the manufacturing process. Injection molding, sheet metal fabrication, and 3D printing are the core commercial services provided by the company.

*FARO Technologies Inc. (FARO)*

FARO specializes in 3D measuring as well as other services for the design, engineering, and construction industries. FARO' has a 40-year history that predates the emergence of 3D printing. Coordinate measuring equipment, laser trackers and projectors, mappers, scanners, and software are among the company's products. FARO also provides services to the aerospace, automotive, and power generating industries.

*Materialise NV (MTLS)*

Materialise, a Belgian firm, has been supplying 3D printing technologies and accompanying software for 30 years. It provides platforms for the development of 3D printing applications in fields such as healthcare, automotive, aerospace, and art & design. Anatomical models in dentistry and hearing aid goods were among the company's earliest 3D printing efforts. Materialise also makes eyeglasses and automobiles.

---

[8] Cf *(5 Biggest 3D Printing Companies, 2022)*

*The ExOne Co. (XONE)*

ExOne specializes in producing 3D printing equipment for customers in a variety of sectors. It also manufactures 3D printed things to order for industrial clients. Binder jetting technology is used by ExOne 3D printers to fuse powder particles of materials such as metal or sand into molds, cores, and other products.

**Table 2: List of 3D printing Companies in descending order**

| 3D Printing Companies | Revenue (TTM) | Net Income (TTM) | Market Cap: | 1-Year Trailing Total Return | Exchange |
|---|---|---|---|---|---|
| 3D Systems Corp. (DDD) | $566.6 million | -$78.4 million | $632.3 million | -24.60% | New York Stock Exchange |
| Proto Labs Inc. (PRLB) | $451.0 million | $58.6 million | $3.9 billion | 58.20% | New York Stock Exchange |
| FARO Technologies Inc. (FARO) | $334.7 million | -$79.7 million | $1.0 billion | 20.60% | NASDAQ |
| Materialise NV (MTLS) | $205.3 million | -$2.7 million | $1.9 billion | 94.80% | NASDAQ |
| The ExOne Co. (XONE) | $52.9 million | -$14.5 million | $238.2 million | 48.30% | NASDAQ |

**Source: (5 Biggest 3D Printing Companies, 2022).**

## 3.4   Future Developments

Which 3D printing materials are predicted to flourish in the coming decade? Will we see freshly found 3D printing materials as a result of AI-enabled computational alloy discovery, with end-users able to define desired properties? What about the widespread use of metamaterials, which are materials having features that do not occur naturally? How will sustainability be considered, such as an increase in attempts to make bio-based polymers or recapture waste plastics and metals, therefore returning resources to a circular economy[9]?

Some of the efforts may be done by the market through end consumers. Purchasing power is used to find successful technological platforms among a sea of imitators. Conversely, as demand for additive manufacturing rises, a swollen market may lift all ships, giving both newcomers and incumbents access to a larger pie to share.

Below is a quote was given by the president and CEO of 3D Systems, Dr. Jeffery Graves, giving an insight into what he thinks the future will be for 3D printing:

*"I expect mass customization will not only be an important trend for 2022 but the coming decade. While many organizations aspire to take advantage of additive manufacturing's ability to produce large quantities of distinct parts, I don't think every organization has fully understood how to integrate AM into its workflow. As we see a broader acceptance of additive alongside traditional technologies, I anticipate we'll also see manufacturers of all sizes embracing AM for mass customization.*

*To facilitate the integration of AM into existing workflows, I believe machine learning will play a critical role. It is not enough to introduce design flexibility, speed to market, or supply chain efficiency offered by additive manufacturing. For companies to maintain their competitive position, they need to have a smart manufacturing strategy to introduce AM*

---

[9] Cf *(Sertoglu et al., 2022)*

*effectively and efficiently into their overall manufacturing workflow. As more companies adopt smart manufacturing solutions, I expect they will see how machine learning can enable autonomous manufacturing – thus helping improve productivity, and enhance capacity to introduce scalability and flexibility into processes.*

*Focusing our perspective on the next decade even further, I expect that we'll continue to see additive manufacturing drive remarkable advancements in the transformation of healthcare delivery. AM has already demonstrated its power in this industry to enable patient-specific healthcare with unique solutions to create surgical plans and medical devices. I'm very excited about the next frontier in healthcare where bioprinting plays an important role. Over the past year, there has been a dramatic increase in the number of entrants to this field, whether it be research organizations or private and public companies. We've seen the scope of the research efforts broaden to include new printing technologies and new materials designed to assist in drug discovery, the creation of tissues, and hopefully one day, producing transplantable human organs. I believe we're on the precipice of amazing advancements in this arena and look forward to what we're able to influence and achieve as an industry."*[10]

In the field of architectural design, there is a significant trade-off between the desire to save money, which entails producing simple and straight walls, and the desire to create novel and nonstandard buildings, which entails producing customized molds and formwork for one-time use, resulting in large amounts of waste materials and unforeseen construction delays[11]. Large-scale 3D printing has the ability to produce complicated geometries that cannot be defined by fundamental geometrical concepts and have previously only been achieved for a few projects. Now that it has been mathematically shown that any 3D building may be formed layer by layer. Structures may be built in practically any shape, regardless of complexity, cost, or potential.

3D printing technology is rapidly evolving, becoming larger, quicker, and more affordable. The spectrum of materials in AM will expand as the need for specialized materials to meet the requisite qualities of end-parts grows. The capacity of the current generation of printers that can handle a wider range of sophisticated materials is critical since it allows businesses to profit from AM where they previously could not. Metal filaments for FDM printing are an excellent example of novel materials[12].

---

[10] *(Sertoglu et al., 2022)*
[11] Cf *(Mourad, Aljassmi and Al Najjar, 2018)*

Although machine prices remain high, part costs are decreasing as speed increases. These developments will increase as more firms use 3D printing. With technologies like dual extrusion, 3D printing's adaptability is increasing and is being seen used in a larger range of sectors. Another noteworthy breakthrough noticed is printing without the usage of support structures, which broadens the spectrum of applications for AM. Could be said, the potential for cost and time savings is quite favorable.

It's not simply about having new and improved printers for manufacturers. Manufacturers would require a wide choice of printers, materials, and, most crucially, contacts with other industry specialists to reap the most benefits. Furthermore, interoperability across all of the various systems is becoming a crucial problem in order to fully utilize 3D printing. Automation in manufacturing and post-processing, as well as integrated usability, will be major trends in 2022 and beyond.

A sustainable manufacturing and supply chain is increasingly important, driven not just by end-customer demand, but also by official requirements and personal convictions. This tendency is also influencing 3D printing. Because 3D printing is an additive manufacturing technique, it may minimize waste during production in the correct application. By precisely designing a part for AM, the weight of the finished item may be dramatically reduced. Furthermore, employing on-demand and decentralized 3D printing might minimize the number of components in inventory and related waste, as well as $CO_2$ emissions during transportation.

As a result, an anticipation increase in the use of 3D printing as part of firms' sustainability strategies in 2022. To boost AM's sustainability, even more, energy usage

---

[12] Cf *(Where is 3D printing heading towards? 6 trends to watch in 2022 - Replique, 2022)*

during production must be decreased. Furthermore, an increase in the use of sustainable AM materials such as recycled, reused, and biodegradable plastics are seen.

## 3.5    Benefits

When the size is no longer an issue, massive pieces may be made at a low cost to replicate enormous goods. Post-processing will make the finished model closer to the original product, and some of the printed pieces will be usable as end-use products (The Ocke Stool, Bathtub after Post Processing, and Children Lamps)[13]. When compared to non-3D printing manual procedures, large-scale 3D printing reduces the time it takes from a concept to a full-size prototype (working with wood, foams, and fiberglass). It also significantly reduces expenditures. Furthermore, the freedom to develop unique shapes and tailor-made goods.

There is also the printing of a single enormous full-scale item or multiple discrete pieces that are joined to form a prototype of a major product or gadget, such as a driver seat or an MRI machine. The advantage is that a prototype of the final part/product may be produced to do fit-testing with other components, design verification, and basic functional testing. When compared to other methods, large-scale 3D printing speeds up design and manufacturing processes while saving money.

This is another example of an industry that has traditionally produced models and things by hand utilizing various processes. The use of a large-scale 3D printer allows for the manufacture of one-of-a-kind marketing and promotional materials. Traditionally, in this sector, all components and models are handcrafted. When compared to traditional techniques, large-scale 3D printing significantly lowers manual labor, saves time, and cuts prices. It also allows for the freedom of design and production of bespoke goods and elements.

The following is a common argument in favor of 3D-printed parts over traditional metal-based ones: You may manufacture components with complicated geometry, porous interiors, lattice structures, and membrane structures using 3D printing; as a consequence, you can make lighter parts with less material, resulting in waste reduction.

---

[13] Cf *(BigRep GmbH, 2017)*

However, this is a naive picture that ignores how 3D printing materials are sourced and what the long-term impacts will be when they become a component of hundreds of thousands of daily things ranging from electronics, wearables, home decor, and furniture to automotive and aircraft parts.

Parts are made from nylon, ABS, thermoplastic polyurethane, and other thermoplastics using the two most prevalent 3D printing processes, FFM and SLS. Technically, pieces made from such materials may be melted and recycled.

# 4   3D Printing Market Analysis

Market participants are constantly improving 3D printing technology in response to the increased demand for 3D printing applications in the automotive, healthcare, aerospace, and military sectors for manufacturing reasons. The leading companies are recognizing business transformation opportunities by using additive manufacturing in new product development processes.

**Table 3: 3D Printing Market Report Scope**

| 3D Printing Market Report Scope | |
|---|---|
| **Report Attribute** | **Details** |
| Market size value in 2022 | USD 16.75 billion |
| Revenue forecast in 2030 | USD 76.17 billion |
| Growth Rate | CAGR of 20.8% from 2022 to 2030 |
| Base year for estimation | 2021 |
| Historical data | 2017 - 2020 |
| Forecast period | 2022 - 2030 |
| Quantitative units | Revenue in USD million/billion and CAGR from 2022 to 2030 |
| Report coverage | Revenue forecast, company ranking, competitive landscape, growth factors, and trends |
| Segments Covered | Component, printer type, technology, software, application, vertical, material, region |
| Regional scope | North America; Europe; Asia Pacific; South America; MEA |
| Country scope | U.S.; Canada; Mexico; U.K.; Germany; France; Italy; Spain; Japan; China; India; South Korea; Singapore; Brazil |
| Key companies profiled | Stratasys, Ltd.; Materialise; EnvisionTec, Inc.; 3D Systems, Inc.; GE Additive; Autodesk Inc.; Made In Space; Canon Inc.; Voxeljet AG |
| Customization scope | Free report customization (equivalent up to 8 analysts working days) with purchase. Addition or alteration to country, regional & segment scope. |
| Pricing and purchase options | Avail customized purchase options to meet your exact research needs. Explore purchase options |

**Source: (3D Printing Market Size & Share Report, 2022-2030, 2022)**

The research anticipates revenue growth at the global, regional, and national levels, as well as an examination of the most recent market trends and prospects in each sub-segment from 2017 to 2030. The research also includes shipping figures and predictions, as well as ASP qualitative analysis from 2017 to 2030. Grand View Research has segmented the worldwide 3D printing market research based on component, printer type, technology, software, application, vertical, material, and region[14].

Spare parts providers are now unable to meet customer demand in the manufacturing and industrial products industries. Industrial products are complicated and comprise a number of distinct parts, the majority of which will need to be changed during the equipment's lifespan. To suit the demands of the consumer, a replacement parts provider is expected to handle a complex network of suppliers, production, sales, and consumers. To keep the spare parts running, several strategic decisions must be made, as well as higher expenditures.[15]

As a result, firms tend to retire an increasing number of parts each year, giving consumers inconvenience. Given the constraints and expense pressures that spare parts buyers face, a growing number of businesses that purchase spare parts are turning to 3D printing to make their own components. As a result, a growing number of customers are turning to 3D printing design and production services.

3D printing suppliers are forming alliances and collaborations with regional players and software vendors to strengthen and collaboratively develop their product portfolios, with the goal of meeting the expanding global demand for sophisticated 3D printing and providing clients with distinctive solutions. In June 2019, for example, the firm announced a collaboration with Siemens, a worldwide technology powerhouse in automation and digitalization. The goal of this collaboration is to connect ExOne's S-Max Pro with Siemens' Digital Enterprise Portfolio of software and automation technology in order to reap the benefits of Industry 4.0.

It also intends to broaden its collaboration with Siemens to include its industry-leading industrial 3D printers for tooling and production metal. Aside from that, in October 2019, 3D Systems announced a cooperation deal with Antleron to offer innovative solutions and speed the development of bioprinting solutions in the biomedical market using 3D Systems' printing technology.

---

[14] Cf *(3D Printing Market Size & Share Report, 2022-2030, 2022)*
[15] Cf *(www.futuremarketinsights.com, n.d.)*

It has been noted that 3D printing solutions and service providers are focused on forming partnerships and collaborations with other regional businesses and raw material providers in the industry to enhance and jointly build their product portfolio throughout the years. These partnerships and collaborations are emerging all the time in order to meet the rising demand for superior 3D printing technologies throughout the world and to offer clients quick access to 3D printing solutions and services.

## 4.1    Market Players

The 3D printing market is concentrated since the bulk of the market share is held by the industry's top players. Small and medium-sized businesses are updating their cloud services, but the top companies have captured a considerable number of users and are spending heavily on new advancements and innovation. Stratasys Ltd.,3D Systems Corporation, EOS GmbH, Electro-Optical Systems, Concept – Laser GmbH, Sisma SpA,

ExOne Co., Arcam AB (GE Aviation), SLM Solutions Group AG, Hewlett Packard Inc., Materialise NV (ADR), and Proto Labs Inc. are only a few of the main companies.

**Figure 1: Major Players**

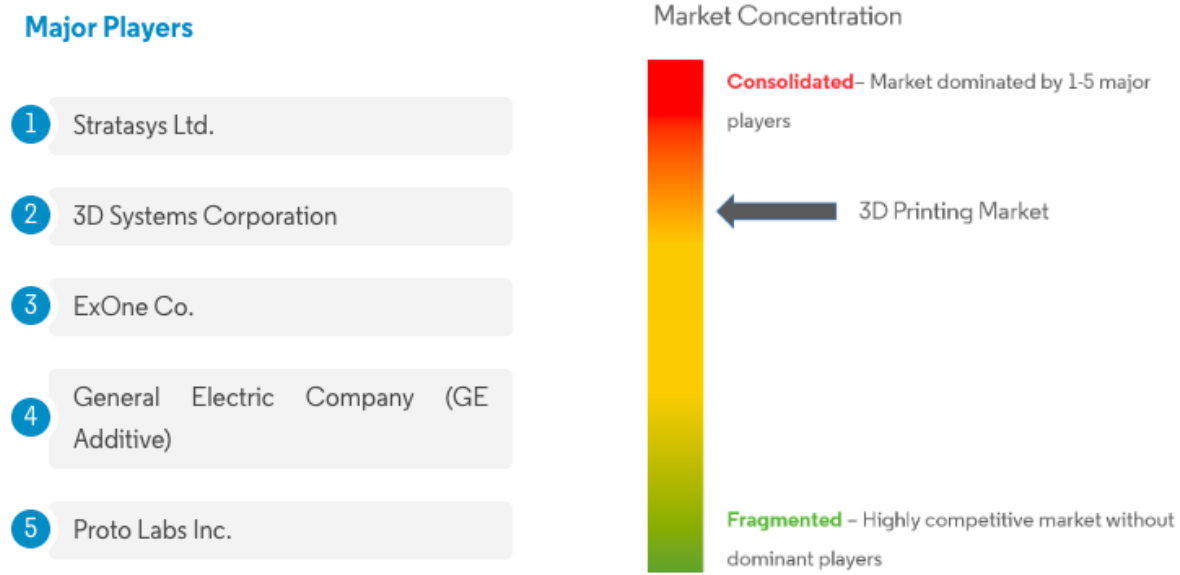

**Source: (3D Printing Market Size, Growth, Trends (2022 - 27) | Industry Forecast, 2022)**

The market's leading competitors are focused on providing improved and new solutions to meet the rising demands of industries. These prominent firms are investing in research and development to create novel services and materials. They are forming strategic alliances and collaborating to develop next-generation solutions. These businesses provide consumer-centric solutions to assist enhance corporate growth. Similarly, the leading companies are eager to provide a diverse selection of 3D materials in order to expand across every industrial application.

In February 2022, for example, Imaginarium teamed with Ultimaker to launch a desktop and industrial 3D printer range in the Indian market. This collaboration will assist

Ultimaker in expanding its company in India, where additive manufacturing is expected to reach a tipping point in the coming years.

3D Systems is then expected to further additive manufacturing innovation through collaborations and product launches in November 2021. The firm introduced new 3D printing tools and partnered with the UK-based startup Additive Manufacturing Technologies to provide unique 3D printing workflows and develop AM software, which is appropriate for automobile components.

## 4.2    Growth Impact made by Market Players due to their Strategies

To increase their market presence, market participants are diversifying their product offerings. Furthermore, developing enterprises are heavily spending on new product development and launches in order to maintain their market position. Furthermore, numerous suppliers are concentrating on increasing the desirability of their solutions among diverse consumers through innovation and development.

Over the projected period, the 3D printing industry is expected to be extremely fragmented. The income generated by large giants and new start-ups, as well as other established small- and medium-sized 3D printing suppliers active in the industry, is attributable to the expected high fragmentation.

Companies with a market share of more than 10% are considered market leaders. This category comprises industry titans like 3D Systems and Stratasys, Ltd., which are the largest and most experienced in the business and have extensive regional presence worldwide.

Companies having a market share of more than 5% but less than 10% are promising. These firms are anticipated to see rapid expansion and capitalize on the possibilities provided by the global market. These firms control around 25-30% of the worldwide market. ExOne, SLM Solutions Group AG, Renishaw plc, and Nanoscribe are among these firms.

Companies with a relatively low share value of less than 5% are attempting to recruit new clients in overseas markets. These firms control between 40 percent to 42 percent of the worldwide 3D printing market.

## 4.3   Largest Global Share in the Market

The North American area is projected to dominate the 3D printing sector, just as it has dominated technological adoption. According to research, the United States is the most advanced country in 3D printing since November 2019.

A slew of new product releases, as well as product improvements and advancements, are likely to boost market growth even further. Several 3D printing solution suppliers are extending their presence in the North American market in order to strengthen their market position. For example, in May 2019, Italian 3D printer maker Roboze debuted its high-temperature ROBOZE Xtreme line in North America for the first time at RAPID+TCT 2019[16].

The area is also seeing a surge in investments in North America's healthcare, aerospace and defense, industrial, and consumer products industries, which are likely to increase considerably in the coming years. Various government agencies, such as NASA, have noticed that significant expenditures on 3D printing technologies may significantly contribute to space applications and the development of zero-G technologies, boosting market growth.

Fitness trackers and smart clothing are also predicted to drive 3D printing technology in the United States. It is expected that around 19% of broadband homes owned a wearable fitness gadget. Furthermore, changing customer tastes and an increasing desire for customization have created a need for a flexible band and electronics systems that might be produced utilizing 3D printing technology, propelling its expansion.

---

[16] Cf *(3D Printing Market Size, Growth, Trends (2022 - 27) | Industry Forecast, 2022)*

**Figure 2: 3D Printing Market - Growth Rate by Region (2021-2026)**

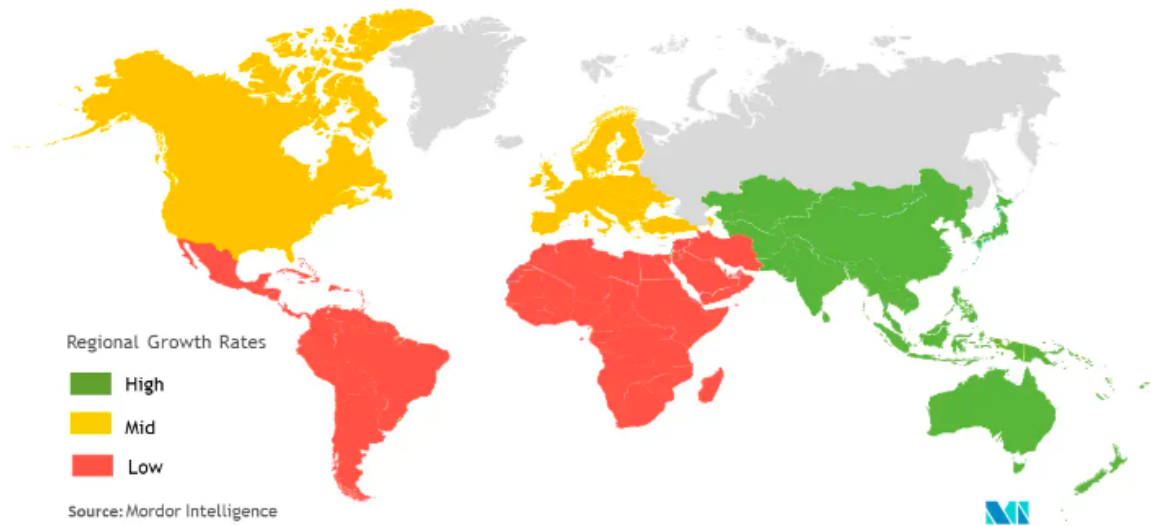

**Source: (3D Printing Market Size, Growth, Trends (2022 - 27) | Industry Forecast, 2022)**

Over the predicted period, Asia Pacific is expected to expand the most past North America. This is due to government-funded research into 3D printed solutions, as well as international investment. This area is predicted to expand significantly due to unmet requirements of a large population base and improved manufacturing infrastructure.[17]

## 4.4 3D Printing in Germany and UK

In the United Kingdom and Germany, the 3D printing business is especially vulnerable to economic swings. In the region, it is largely employed as a prototyping process.

---

[17] Cf *(Research, 2022)*

Over the previous two decades, 3D printing industry participants have lowered R&D spending, resulting in a decline in demand for prototype services.

By bringing 3D printing processes in-house, the production time for 3D printing manufacturing processes may be cut to days, if not hours. As a consequence, the whole product design and production cycle are reduced, and repetitive redesigning, as well as lengthier wait periods, are avoided. As a result, the aerospace and defense industries are expected to continue to see increased use of 3D printing software and solutions in Europe throughout the projection period. Bringing 3D printing techniques in-house, on the other hand, maybe difficult and costly. As a result, not all businesses can afford to manufacture their own 3D printed items.

## 4.5    3D Printing Market Size Forecast with COVID-19 Impact

The introduction of COVID-19 has had a significant impact on the 3D printing sector, owing to a scarcity of competent individuals to operate the technology. Furthermore, the pandemic had an influence on the economy and the operation of manufacturing sectors, which considerably stimulated market growth. In contrast, the market had a revenue impact as a result of worldwide manufacturing unit shutdowns[18].

The COVID-19 pandemic has disrupted the ecosystem's supply chains. In terms of the market, the pandemic has had a wide-ranging impact on industries such as healthcare, automotive, aerospace, consumer electronics, retail, energy and power, oil & gas, construction, jewelry, food & culinary, and education, to name a few. During the pandemic, the healthcare industry experienced an unprecedented surge in demand for personal protection equipment including face masks, shields, and ear bands. Furthermore, the demand for venturi valves and regulators that help patients breathe has increased. Aside from that, e-commerce has thrived due to regional lockdowns and logistical difficulties. 3D printing firms may directly print and distribute items and have total control over the materials used in manufacturing, storage conditions, and distribution. They can also offer digital files that allow customers to print their own devices.

---

[18] Cf *(3D Printing Market Size, Share | Trends & Forecast - 2030, 2022)*

The COVID-19 pandemic has also hampered multiple enterprises in a variety of industries, including automotive, aerospace and military, consumer electronics, food and drinks, and retail. Import and export restrictions from China and other APAC nations have prevented industrial facilities in North America and Europe from operating[19]. This has resulted in considerable fluctuations in overhead costs and overall product quality for these producers. The International Monetary Fund (IMF) forecast a more than 8% drop in GDP owing to the coronavirus pandemic in April 2020. One of the leading manufacturers of Binder Jetting systems reported a huge drop of 27% in the second quarter of 2020.

## 4.6   3D Printing Application Analysis

**Figure 3: Global 3D Printing Market Share - by Application 2021**

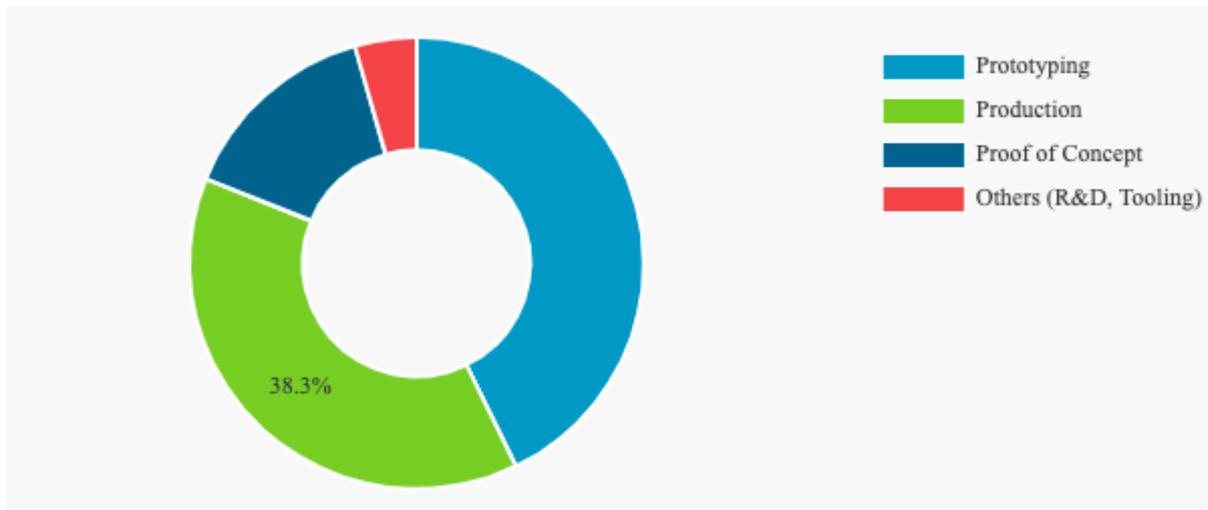

**Source: (www.fortunebusinessinsights.com, n.d.)**

Prototype held a substantial market share in 2021 due to the broad acceptance of the prototyping process across many vertical sectors. Prototyping enables firms to attain higher precision and consistently generate finished goods. This technology aids in the production of 3D CAD models and prototypes.

---

[19] Cf *(Market, 2022)*

During the projection period, the output is expected to expand rapidly. While businesses move their traditional production units to advanced manufacturing processes, the use of this technology to produce complicated and low-volume products is likely to increase throughout the projection period.[20]

## 4.7    Rise in Investment from Governmental bodies in 3D Printing Projects

Because 3D printers are considered a growing global market, government organizations all around the world are backing 3D printer producers with specific policies to help them overcome financial difficulties. In addition, the government is offering specific educational workshops to help people overcome technical and economic difficulties. These authorities' policies contribute to a healthy environment for the design, development, and deployment of 3D printers. 3D printing, also known as additive manufacturing, is a method of creating prototypes or working models of products by layering materials such as plastic, resin, thermoplastic, metal, fiber, or ceramic. The model to be printed is created by the computer using software, which then sends instructions to the 3D printer. The main problem that producers may encounter is the availability of raw materials, as it is regarded as a highly specialized industry.

## 4.8    Things hindering the 3D Printing Market Growth

One of the primary issues now confronting 3D printer producers, particularly those in developing nations, is a scarcity of competent individuals. The cost of raw materials is another aspect that is predicted to stymie the industry, as the bulk of materials is created by patent-holding corporations. As a result, nations like Brazil, India, Australia, and others may encounter cost issues. These are some of the issues that the Global 3D Printer Market is now facing.

---

[20] Cf *(www.fortunebusinessinsights.com, n.d.)*

## 4.9   Market Opportunities

The ongoing transition to industrial automation will provide future prospects for macro 3D printers. For example, the future of the construction sector will feature 3D printers, drones, and robotic bulldozers, which will aid in the creation of robotic construction sites. Furthermore, Local Motors, a US-based business, has already produced Strati, a 3D-printed automobile.

Spark, an open collaboration platform for 3D printing, assists firms in developing various types of 3D printing technologies by offering extensible Application Programming Interface (API) for all phases of the 3D printing workflow. Companies may use this program to create 3D models for any 3D printer. This sort of platform is assisting in raising awareness and promoting macro 3D printing.[21]

---

[21] Cf (Persistence Market Research, n.d.)

# 5 Final Consideration

## 5.1 Results and critical reflections

With the aid of 3D printing, the world is constantly evolving. The usage of 3D printing for therapeutic purposes today is astounding, but what the future holds is uncertain; yet, additive layer manufacturing will play a significant role in fixing our issues. 3D printing truly is boundless, and we have just scratched the surface; there is much more to be discovered. As seen across the website. Although 3D printing bones is still in its early stages and is constantly being improved and adjusted, it has already improved the lives of many people throughout the world, particularly in Australia. It is obvious that the more money and research put into 3D printing, the further it will lead us. 3D is inherently unexpected.[22]

The year 2022 appears to be "The year of 3D Printing". A slew of new discoveries will propel AM industrialization ahead. Printing times are very long, which is one of the key reasons why AM has not yet achieved large-scale manufacturing. Injection molding enables the production of standardized components in a consistent and timely manner. However, 3D printing is still a relatively young technology that is rapidly evolving. The cost of such printers is falling as manufacturing speed, quality, and build plate size improve. 3D printing will also enable a more resilient and sustainable supply chain strategy, making the technology more appealing to businesses. Security, quality assurance, and interoperability advancements will make the technology more accessible to a broader spectrum of companies. Even though we are still a long way from mass manufacturing, we anticipate potential production quantities for 3D printing increasing.

It is widely assumed that 3D printing will be a transformative force in manufacturing, whether positive or negative. Despite worries about counterfeiting, numerous firms are already employing the technology to make sophisticated components in a repeatable manner, such as in automotive and aerospace production.[23]

---

[22] Cf (3d printing is limitless, n.d.)

As 3D printers grow more inexpensive, they will eventually be employed for local, small-scale production, obviating the need for many forms of supply networks. Consumer units for home usage will even be possible, allowing end customers to simply download and print a design for the product they desire.

The traditional manufacturing industry will have significant hurdles in adapting to these developments. However, the prospects for technology and engineering are certainly enormous, as are the creative possibilities in product design and printing material formulation.

Recent scientific breakthroughs and uses of 3D printing indicate that the technology has the potential to transform many aspects of daily life. AM's influence on supply chains, for example, takes various forms, including simpler production processes, decreased material waste for leaner manufacturing, improved flexibility, lower prices, faster response to demand, and the capacity to decentralize production.[24]

## 5.2    Implications for further research

For further research, 3D printing technology must first overcome a major challenge. There are still perspectives on 3D printing from its beginnings more than 5 years ago to supply more knowledge among people about all of its capabilities and advancements. Today, the reality is different, and the argument that 3D printing increases raw material use could be deceptive. The exact opposite is said to be true. 3D printing could have a significant influence on trash logistics. However, favorably, a well-designed 3D printing production process leads to considerable reductions in resource consumption and waste generation; because manufacturing will be closer to the consumer, packing and transportation materials will be used less often than in the past. The basic nature of the additive process results in reduced waste in production.

---

[23] Cf *(AZoM.com, 2012)*
[24] Cf  *(Kubáč and Kodym, 2017)*

Another incorrect perception of 3D printing is that its primary application is for plastic things, generally as components to finished products. That is not correct. The key advantage of the technology is the ability to employ a variety of materials other than plastics. Metal 3D Printing, for example, has the capability and capacity to build delicate, streamlined components with physical qualities that can sometimes exceed those of parts made traditionally. As a result, the technology has the potential to totally transform the way we manufacture crucial components. It could be used to make lightweight items with distinctive shapes that reduce material waste and energy usage.

In other words, there is a "knowledge gap" between the existing technology and the people that must be bridged before 3D printing can become a mainstream technique of product creation.

## 5.3   Implications for practice

Not all 3D printers are the same. They vary depending on the application. They differ depending on the material. They differ in terms of output and performance. As a result, while embarking on a 3D Printing journey, it is critical to have a clear grasp of what your supply chain looks like, what the customer's expectations are, and where it receives its information from in order to pick the appropriate equipment and procedures.

Discussions about design, iteration, production, warehousing, shipping, warehousing, and distribution to customers in a classic manufacturing setting and supply chain have been done. The 3D printed supply chain has the potential to be considerably shorter, more efficient, and tailored to the needs of the consumer. It reduces the design cycle, optimizes the iteration cycle, improves the manufacturing cycle by generating unique items as needed, eliminates the warehouse, and transforms the distribution routes into a 3D Printed Supply Chain. This often necessitates a full rethinking of your business as well as a possibly more collaborative engagement with your consumer. In fact, a company model may alter so drastically that it begins licensing the intellectual property or data, allowing the client to manufacture the goods on their premises.

The ramifications are enormous. If a mere examination has trainers, sneakers, and shoes, have an influence on the materials produced by major chemical plants, logistics, the way money is exchanged, the retail environment, and the size and type of the factories. Changes the labor component, possibly altering the dynamics and ethical sourcing problems. Trade and tariffs have macroeconomic implications, as does the collecting of duty money for national governments.[25]

For further reading of the Logistics Engineering and Technologies Group please refer to Auerbach & Uygun, 2007; Keßler et al., 2007; Keßler & Uygun, 2007; Kortmann & Uygun, 2007; Droste et al., 2008; Uygun, 2008; Kuhn et al., 2009; Uygun & Wötzel, 2009, Jungmann & Uygun, 2010; Keßler & Uygun, 2010; Uygun & Kuhn, 2010; Uygun & Luft 2010; Uygun & Schmidt, 2011; Uygun & Wagner, 2011; Liesebach et al., 2012; Uygun et al., 2012; Uygun, 2012a; Uygun, 2012b; Uygun, 2012c; Uygun & Straub, 2012, Besenfelder et al, 2013a; Besenfelder et al 2013b; Güller et al., 2013; Scholz et al., 2013; Uygun, 2013; Uygun & Straub, 2013; Güller et al., 2015; Mevenkamp et al., 2015; Uygun et al., 2015; Karakaya et al., 2016; Uygun & Reynolds, 2016; Güller et al., 2017; Reynolds, & Uygun, 2018; Uygun & Ilie, 2018; Lyutov et al., 2019; Nosheen & Uygun, 2020; Sommerfeld & Uygun, 2020; Uygun & Jafri, 2020; Uygun, 2020; Özgür et al., 2021; Lyutov et al., 2021; Uygun & Ahsan, 2021; Uygun & Rustemaj, 2021; Uygun et al., 2022, Merten et al., 2022a, Merten et al., 2022b, Lyutov et al., 2022.

---

[25] Cf *(www.linkedin.com, n.d.)*

# Table of references